\documentclass[a4paper,11pt]{article}
\usepackage{pos}

\title{Low energy constant and mixed-action effect}

\author*[a,b]{Dian-Jun Zhao}
\author[a,b,c,d]{Yi-Bo Yang}

\affiliation[a]{School of Physics Sciences, University of Chinese Academy of Science,\\
100190, Beijing, China}
\affiliation[b]{CAS Key Laboratory of Theoretical Physics, Institute of Theoretical Physics, Chinese Academy of Sciences\\
100190, Beijing, China}
\affiliation[c]{School of Fundamental Physics and Mathematical Sciences, Hangzhou Institute for Advanced Study, UCAS\\
310024, Hangzhou, China}
\affiliation[d]{Internationcal Centre for Theoretical Physics Asia-Pacific\\
Beijing/Hangzhou, China
\vspace*{-0.5cm}
\begin{center}
\large{
\vspace*{0.4cm}
\vspace*{0.4cm}
($\chi$QCD Collaboration)}
\end{center}}
\emailAdd{zhaodianjun@itp.ac.cn}
\emailAdd{ybyang@mail.itp.ac.cn}

\abstract{We present the pion mass and decay constant using the overlap fermion valence on domain wall (DW) fermion sea at several lattice spacings. The mixed action effect in the lattice calculation is also studied, and the result suggests that the mixed action effect with the overlap valence on DW sea would be proportional to the fourth power of the lattice spacing.  We obtain the pion decay constant at the physical pion mass and $N_f=2$ chiral limit to be $92.4(3)(2)~{\rm MeV}$ and $87.0(5)(7)~{\rm MeV}$ respectively; and the physical u/d averaged quark mass at $\overline{\textrm{MS}}$ 2 GeV is $3.74(4)(5)(5)(3)~{\rm MeV}$ with the linear ${\cal O}(a^2)$ continuum extrapolation. Using the FLAG value of the NLO low energy constant, we obtain the $N_f=2$ chiral condensate to be $\Sigma^{\overline{\textrm{MS}}(2~{\rm GeV})}=\big(266(2)(1)~{\rm MeV}\big)^3$. }

\FullConference{
 The 38th International Symposium on Lattice Field Theory, LATTICE2021
  26th-30th July, 2021
  Zoom/Gather@Massachusetts Institute of Technology
}

\begin{document}
\maketitle

\section{Introduction}
The low energy feature of the strong interaction binds the quarks into the mesons and hadrons, and creates most of the mass of the visible matter. At the leading order of the chiral perturbative theory, there are just two additional parameters beside the quark mass: the chiral condensate and pion decay constant. Since those values are quark mass dependent, it is essential to determine them from the first principle lattice QCD calculation. 

In this proceeding, we provide the preliminary lattice QCD result of the quark mass and decay constant at the physical pion mass, and also the leading order (LO) low energy constants.

\section{Numerical setup and the mixed action effect}

The basic information of the RBC/UKQCD 2+1 flaovr dynomcial ensembles we used in this project is summarized in Table \ref{Tab1}, which “DW” refers to domain wall fermion action, while “I” indicates that the gauge field is Iwasaki. For more information about these two ensembles, see 
\cite{RBC:2014ntl}. We used the overlap fermion action~\cite{Chiu:1998gp} for the valence quark, with 1-step HYP smearing on the gauge field and $\rho=1.5$.

\begin{table}[h]      
\centering
\caption{Basic information of the ensembles we used.}
\label{Tab1}
    \begin{tabular}{ccccc}    
        Action & Symbol & $L^3\times T$ & $a({\rm fm})$ & $m_{\pi,ss}({\rm MeV})$ \\   
        \hline
        Overlap & 24J & $24^3\times\ 48$ & $0.112$ & $290$ \\
        \hline
        DW+I & 48I & $48^3\times\ 96$ & $0.114$ & $139$ \\
        \hline
        DW+I & 64I & $64^3\times 128$ & $0.084$ & $139$ \\
        \hline       
        \end{tabular}
\end{table}

Using different valence and sea actions can introduce the mixed action effect. For example, if we calculate the valence quark propagator using discretized fermion action A on a gauge ensemble with dynamical light quark using another action B, and tune the valence quark mass to make the corresponding pion mass $m_{\pi,vv} = m_{\pi,ss}$, then the pion mass $m_{\pi,vs}$ of the pion correlator formed by one valence quark with one action and one sea (anti-)quark with another action, can suffer from a discretization error and larger than $m_{\pi,vv}$.

\begin{figure}[h]
\centering
\includegraphics[width=10cm]{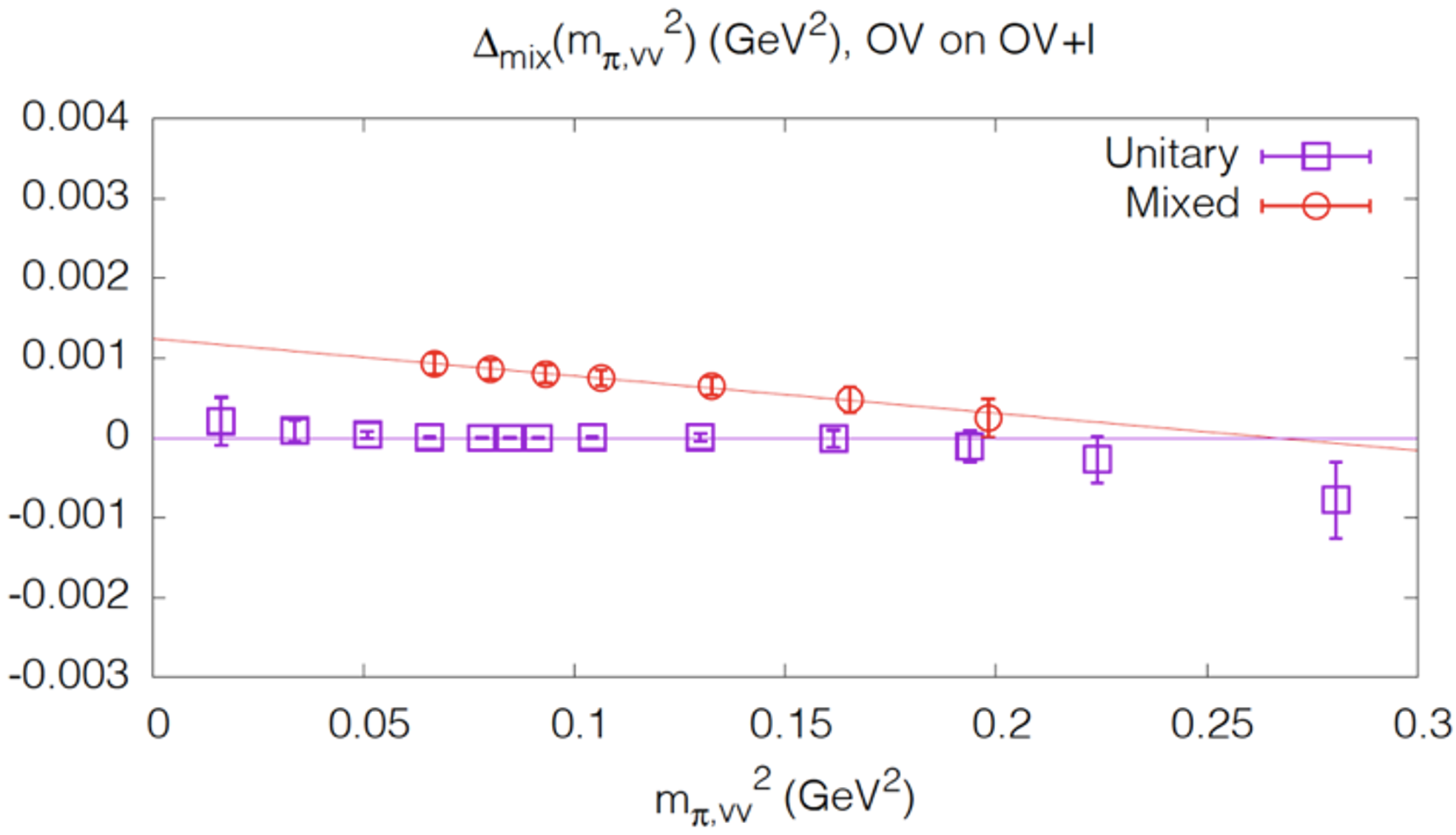}
\caption{Mixed action effect on the JLQCD ensemble with $m_{\pi}\sim$ 290 MeV and $a\sim 0.11$ fm~\cite{Fukaya:2010na}. The red dots and purple boxes show the cases with and without HYP smearing on the valence quark action.}\label{fig:mixed_jlqcd}
\end{figure}

Fig.~\ref{fig:mixed_jlqcd} shows the mixed action effect defined by 
\begin{equation}
    \Delta_{mix}=m_{\pi,vs}^2-\frac{m_{\pi,vv}^2+m_{\pi,ss}^2}{2},
\end{equation}
on the JLQCD ensemble at 0.11 fm with $\sim$ 300 MeV pion mass. The purple data points in the figure show the case with the same valence and sea actions (No HYP smearing, $\rho=1.3$), and $\Delta_{mix}$ is consistent with zero within the statistical uncertainty which is relatively larger when the valence pion mass is far away from the sea one. But if we apply the HYP smearing in the definition of the valence quark action, $\Delta_{mix}$ becomes obvious non-zero, as the red data points in Fig.~\ref{fig:mixed_jlqcd}. It suggests that $\Delta_{mix}$ can actually provide a good reference on the mixed action effect.

\begin{figure}[h]
\centering
\includegraphics[width=10cm]{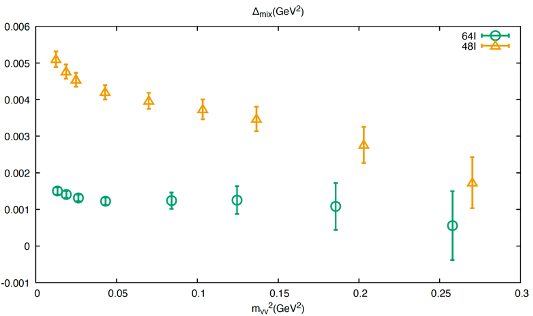}
\caption{Mixed action effect on the RBC/UKQCD ensembles with $m_{\pi}\sim$ 139 MeV and $a\sim 0.08$ fm (green dots) and 0.11 fm (orange triangles) respectively.}\label{fig:mixed_rbc} 
\end{figure}

On the RBC/UKQCD ensembles at physical pion mass, the mixed action effect is much larger, while it is still much smaller than our previous estimate~\cite{Lujan:2012wg} which is around 0.01 GeV$^2$ at a=0.111 fm. At the same time, the values at $a\sim 0.08$ fm are much smaller than those at a$\sim$0.11 fm, and the lattice spacing dependence seems to be ${\cal O}(a^4)$ instead of the widely used ${\cal O}(a^2)$ estimate. 

But even though the mixed action effect here has been much smaller than those in the literature with the other setups (likes the Overlap on Clover~\cite{Durr:2007ef}, DW on staggered~\cite{Orginos:2007tw,Aubin:2008wk}, or mobius DW on the gradient-flowed HISQ~\cite{Berkowitz:2017opd}), it is still much larger than the statistical uncertainty and then the mixed action pion mass can not be used in the analysis. Thus we will just consider the valence pion mass around the unitary point to extract the low energy constants. 

\section{Results}

In this proceeding, we use the low mode substitution (LMS) to obtain good signal with reasonable cost \cite{xQCD:2010pnl}. More precisely, we replace the low mode part of the pion correlation function with small statistics by that from the the all-to-all propagator using the low lying eigenvectors,
\begin{equation}
    C_2^{\rm LMS}(t_f)=\frac{1}{N}\sum_{i}[C_2(t_f,\vec{x}_i,0;S)-C_2(t_f,\vec{x}_i,0,S_{\rm low})]+\frac{1}{L^3\times T}\sum_{t_0,\vec{x}}C_2(t_f+t_0,\vec{x},t_0;S_{\rm low}),
\end{equation}
where
\begin{equation}
    \langle C_2(t_f,\vec{x},t_0;S)\rangle=\langle\sum_{\vec{y}}Tr[S^{\dagger}(\vec{y},t_f;\vec{x},t_0) S(\vec{y},t_f;\vec{x},t_0)]\rangle,
\end{equation}
$S_{\rm low}$ is the low mode part of the point source (or grid source without gauge fixing, equivalently) propagator $S$, and $x_{1,...,N}$ are $N$ kinds of the source positions used in the production to estimate the rest part of $C_2$.

\begin{figure}[h]
\centering
\includegraphics[width=15cm]{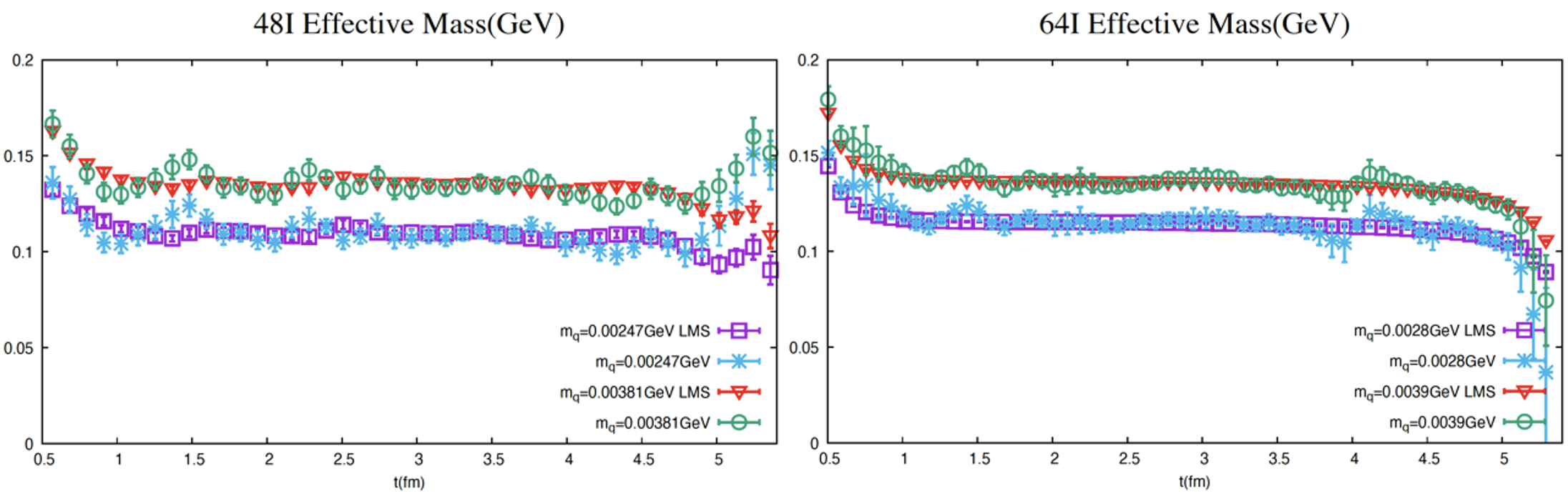}
\caption{The effective mass of the pion correlation function around the physical point, on two ensembles at 0.11 fm (left panel) and 0.08 fm (right panel). The cases with LMS have much better signal and the statistical fluctuations are significantly suppressed.}\label{fig:eff_mass}
\end{figure}

As in Fig.~\ref{fig:eff_mass}, the effective decay rates of $C_2$ (green dots) show obvious statistical fluctuation as functions of the source-sink separation $t$ if the LMS is not applied. But when its low mode part is replaced by the all-to-all propagators, the fluctuation is highly suppressed and then one can extract the pion mass with much higher precision. Thus based on the linear interpolation, we obtained the renormalized quark mass at $\overline{\textrm{MS}}$ 2 GeV with $m_{\pi}=134.98$ MeV as 
\begin{equation}
        m^{\overline{\textrm{MS}}(2~{\rm GeV})}_{ud}(0.114~\mathrm{fm})=3.80(4)(10)(3)~{\rm MeV},\ 
        m^{\overline{\textrm{MS}}(2~{\rm GeV})}_{ud}(0.084~\mathrm{fm})=3.77(1)(7)(2)~{\rm MeV},
\end{equation}
at two lattice spacings respectively, and three uncertainties are the statistical error, the truncation error of the perturbative matching from the RI/MOM scheme to the MS-bar scheme, and the other uncertainty of the renormalization constants $Z_S$~\cite{He:2021tve}. After a linear $a^2$ extrapolation, we predict the average light quark mass to be
\begin{equation}
        m^{\overline{\textrm{MS}}(2~{\rm GeV})}_{ud}=3.74(4)(5)(5)(3)~{\rm MeV},
\end{equation}
where we take the difference between the values at the smaller lattice spacing and that in the continuum as the last uncertainty, to estimate the systematic uncertainty from the continuum extrapolation. Note that the truncation error of the perturbative matching at different lattice spacing are correlated and then will be suppressed during the continuum extrapolation. Such a value is much larger than the current FLAG average, 3.36(4) MeV. It would related to the RI/MOM renormalization scheme we used, as we found that the SMOM scheme used by part of the quark mass and condensate calculations would be have additional systematic uncertainty due to its highly non-trivial $a^2p^2$ dependence~\cite{He:2021tve}. 

\begin{figure}[h]
\centering
\includegraphics[width=15cm]{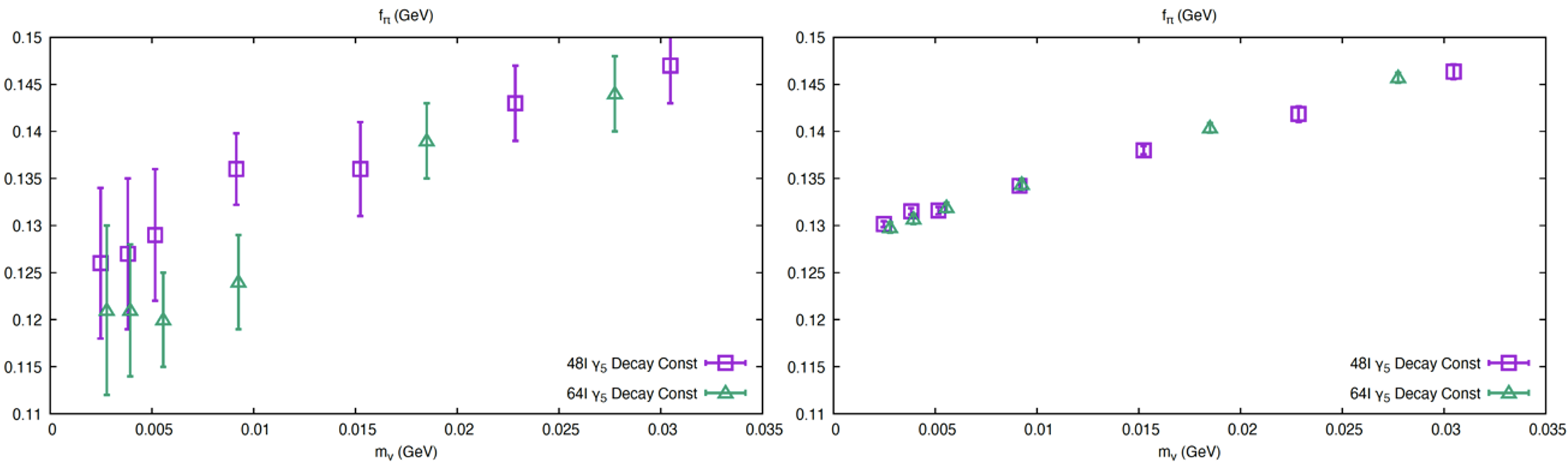}
\caption{The pion decay constant without (left panel) and with (right panel) LMS at $a\sim 0.11$ fm (pruple boxes) and 0.08 fm (green triangles).}\label{fig:decay_constant}
\end{figure}

For the pion decay constant, one can obtain it from the correlation function with either the $\bar{\psi}\gamma_5\gamma_4\psi$ or $\bar{\psi}\gamma_5\psi$ interpolation field,
\begin{align}
    (m_{q_1}+m_{q_2})\langle 0|\bar\psi \gamma_5\psi|\pi\rangle&=M^2_{\pi}f_{\pi},\nonumber\\
    Z_A\langle 0|\bar\psi \gamma_4\gamma_5\psi|\pi\rangle&=M_{\pi}f_{\pi}.
\end{align}
Since the latter definition will depend on the value of $Z_A$, we will concentrate on the first definition to extract the pion decay constant. The result without and with LMS at two lattice spacings are shown in Fig.~\ref{fig:decay_constant}. One can see that the LMS is very helpful to improve the signal, and the results at two lattice spacings are consistent with each other. Eventually we obtain the pion decay constant at the physical point as
\begin{equation}
        f_\pi(0.114~\mathrm{fm})=0.1301(3)~{\rm GeV},\ 
        f_\pi(0.084~\mathrm{fm})=0.1304(1)~{\rm GeV},
\end{equation}
and that in the continuum with the linear $a^2$ extrapolation is
\begin{equation}
        f_\pi=0.1307(4)(3)~{\rm GeV},
\end{equation}
where the second error estimates the systematic uncertainty from the continuum extrapolation by taking difference between the value at smaller lattice spacing and continuum.

\begin{figure}[h]
\centering
\includegraphics[width=14cm]{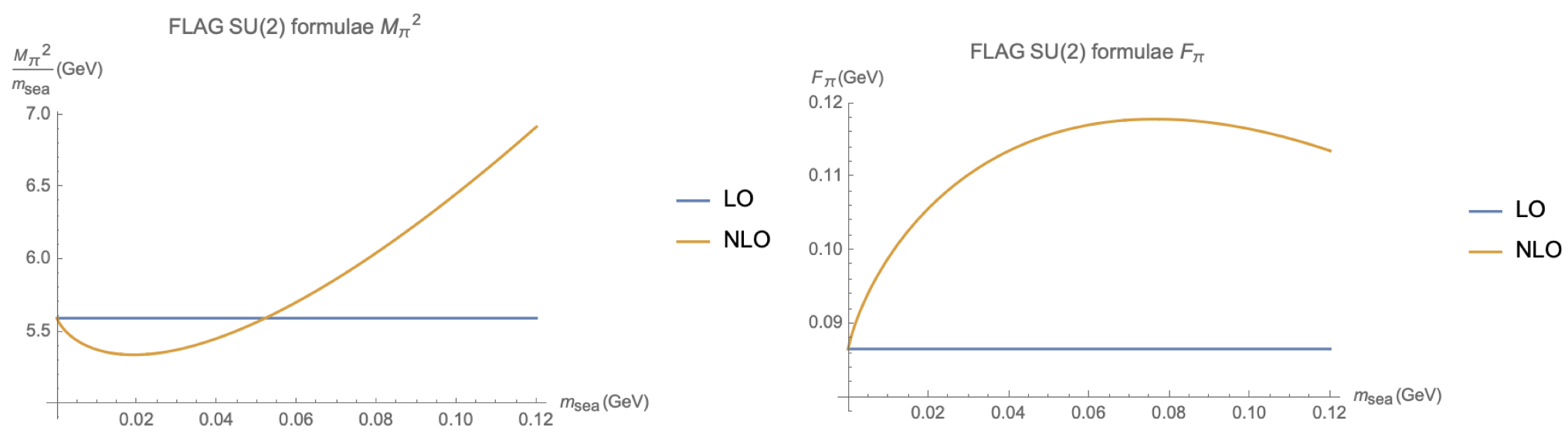}
\caption{Quark mass dependence of the pion mass and decay constant based on the FLAG values.}\label{fig:flag}
\end{figure}

The expansions of the pion mass square and decay constant as the function of the low energy constants $\Sigma$, $F$ and also quark mass $m$ are known based on $N_f=2$ chiral effective theory (Note that $F_{\pi}\equiv f_{\pi}/\sqrt{2}$),

\begin{align}
    M_\pi^2&=M^2[1-\frac{1}{2}x(\ln\frac{M^2_{\pi, {\rm phys}}}{M^2}+{\ell}_3)+\mathcal{O}(x^2)],\\
    F_\pi&=F[1+x(\ln\frac{M^2_{\pi, {\rm phys}}}{M^2}+\ell_4)+\mathcal{O}(x^2)],
\end{align}
where $x=\frac{M^2}{(4\pi F)^2}$, $M^2=\frac{2\Sigma m}{F^2}\propto m$, $M_{\pi, {\rm phys}}=134.98$ MeV, and the lattice averages of the next leading order (NLO) low energy constants are ${\ell}_{3}=3.07(64)$ and ${\ell}_4=4.02(45)$  respectively ~\cite{FlavourLatticeAveragingGroup:2019iem}. Based on the lattice averages of $\Sigma$ and $F$, the quark mass dependence of the pion mass and decay constant are shown in Fig.~\ref{fig:flag}.

If we use the FLAG values of ${\ell}_{3,4}$ and our determination of the physical quark mass and decay constant, we would estimate 
\begin{align}
F=87.0(5)(7)~{\rm MeV},\quad \Sigma^{\overline{\textrm{MS}}(2~{\rm GeV})}&=\big(266(3)(1)~{\rm MeV}\big)^3.
\end{align}

\section{Summary}
In summary, we calculated the pion masses and decay constants with different valence quark masses using the overlap fermion, on the 2+1 flavor DW ensembles with physical pion mass at two lattice spacings. We found that the mixed action effect is much smaller than the other mixed action setup so far, while it is still much larger than the statistical uncertainty needed by a high precision quark mass determination. Thus we only consider the pure valence pion mass at the unitary point, and use the current FLAG averages of the NLO low energy constants to determine the LO low energy constants, $\Sigma$ and $F$. The preliminary estimates are the following: 
\begin{align}
    m^{\overline{\textrm{MS}}(2~{\rm GeV})}_{ud}&=3.74(4)(5)(5)(3)~{\rm MeV},\nonumber\\
    \Sigma^{\overline{\textrm{MS}}(2~{\rm GeV})}&=\big(266(2)(1)~{\rm MeV}\big)^3,\nonumber\\
    F_\pi&=92.4(3)(2)~{\rm MeV}, \nonumber\\
    F&=87.0(5)(7)~{\rm MeV}.
\end{align}

The decay constant we obtain here is consistent with the FLAG value 86.2(5)~MeV. But the chiral condensate $\big(266(2)(1)~{\rm MeV}\big)^3$ is somehow lower than the FLAG average $\big(272(5)~{\rm MeV}\big)^3$, while close to our previous determination from the Dirac spectrum~\cite{Liang:2021pql}, $\big(260(1)(2)~{\rm MeV}\big)^3$. The tension between those two values is under investigation. If we force the physical quark mass to be the FLAG average 3.36(4) MeV, then the chiral condensate will be $\big(276(3)~{\rm MeV}\big)^3$ and perfectly consistent with the FLAG average of the chiral condensate. 

In the next step, we would try to combine the results of the other ensembles to extract the LO and NLO low energy constant simultaneously, it would suppress the uncertainty of our determination of $F$ and $\Sigma$. At the same time, the other sources of the systematic uncertainties like the lattice spacing determination, isospin symmetry breaking effect, partially quenching effect (the sea pion mass is 4 MeV higher than 135 MeV) and so on, should be investigated.  

\bibliographystyle{unsrt}
\bibliography{ref}

\end{document}